# ASI Space Science Data Center participation to high-school outreach program

Angelo Zinzi*[1,2], Carlotta Pittori*[1,3], Rosa Tagliamonte[2], Elisa Nichelli[1,3]

1) Space Science Data Center – ASI, Via del Politecnico, snc, 00133, Rome, Italy
2) Agenzia Spaziale Italiana, Via del Politecnico, snc, 00133, Rome, Italy
3) INAF – Osservatorio Astronomico di Roma, Via Frascati 33, 00078, Monte Porzio Catone, Italy

*Corresponding authors:
Angelo Zinzi, angelo.zinzi@ssdc.asi.it, +39 068567884
Carlotta Pittori, carlotta.pittori@ssdc.asi.it, +39 068567469

# Abstract

Since 2017 the Italian Space Agency (ASI) participates to so-called "Alternanza Scuola-Lavoro" (i.e., "school-work synergy") outreach projects promoted by the Italian government, and the ASI Space Science Data Center (SSDC) actively contributes to them, with the primary aim of bringing students closer to space-related activities before choosing their university studies.

The SSDC outreach program is split into two parts: one theoretical, in which relevant topic are presented and explained, and one practical, consisting of hands-on activities aimed to replicate scientific analysis of real space data. The impact of the program on students' attitude is then evaluated by means of questionnaires specifically designed to gather information on the students' educational background, the level of engagement triggered by the proposed activities, their relevance to school-based activities, and the perceived ease of understanding of the covered topics. As reported in this paper, the analysis of the answers clearly shows that students greatly appreciated this outreach project, supporting its possible expansion and development, even articulated in a more complex pedagogical plan, as already done for one school in a pilot case. Therefore, we plan to expand these activities in the next future both by including new topics (e.g., cosmology, stellar physics), and by proposing new more articulated teaching pathways inclusive of on-site activities in the classroom.

# 1. Introduction

The world we are currently living in offers a wide range of opportunities to keep in touch with scientific and technological realities. Nonetheless, distrust toward scientific themes and actors seems to be spreading among the population, as demonstrated by international surveys highlighting a decline of interest for science studies and mathematics in young European people (European Commission, 2007; Tagliamonte, 2017, OECD, 2003; OECD, 2014; OECD 2016). To reverse this trend, the international science education community mostly agrees on the importance to adopt pedagogical practices on inquiry-based methods, emphasizing the student's role in the learning process, and several projects already focus on this issue (e.g., Center for Informal Learning in Schools - https://www.exploratorium.edu/education/center-informal-learning-schools; European Space Education Resource Office - http://www.esa.int/Education/Teachers_Corner/About_the_ESERO_project; NASA STEM: Explore NASA STEM https://www.nasa.gov/stem). A motivating context for teaching and learning STEM (Science, Technology, Engineering, Mathematics) is important to engage students in their studies, and to pursue a career in these fields, see for example (European Commission, 2015 and references therein) or (Kennedy and Odell 2014).

The goal of this paper is to illustrate how students involved in pedagogical programs designed by the Italian Space Agency (ASI – Agenzia Spaziale Italiana), in particular describing the activities proposed by the ASI Space Science Data Center (SSDC), reacted to the experience, thus trying to provide a useful example for similar projects that could be proposed by other scientific organizations in the future.

Indeed, thanks to both the attractiveness of space programs in the collective imaginary and their strict link with several STEM topics (Quinn, Schweingruber, Feder 2008), the potential of using space as a teaching and learning context for STEM subjects in primary and secondary schools is exploited with success by many space agencies all over the world (ESA 2015; NASA's Education Programme - http://www.nasa.gov/offices/education/about/index.).

ASI has always been a reference point in Italy for scientific training, dissemination of aerospace culture, and promotion of educational initiatives and projects for schools. In particular, for high-school students, ASI offers educational programs whose main goal is to introduce students to activities that will take place in research institutes, aerospace industries, and universities.

Special care is also given to the relationship with teachers, encouraging training in space themes, providing teaching support materials, and organising dedicated conferences and workshops. The goal is enabling capacity building in the teacher communities, and designing and promoting the use of resources that make use of space as a context for the teaching and learning of STEM-related disciplines.

Along these lines, the "Alternanza scuola-lavoro" (i.e., "School-work synergy") project promoted by the Italian Ministry for the Education, University and Research (MIUR) allows high-school students to effectively try different work experiences, in order to get basic skills to smoothly access the labour market at the end of their scholastic path.

ASI proposed a program for this project during the school years 2017/'18 and 2018/'19, and two different schools have been selected each year among a list of candidates. The ASI SSDC actively contributed to the programs, letting the students work on real scientific space data and tools developed for professional astronomers. To understand the impact of these SSDC programs, specifically designed questionnaires have been submitted to the students at the end of the "hands-on" SSDC activities. As described in the following sections of this paper, the students' answers allowed us to evaluate how much the proposed experience was attractive to them, and how it influenced their view of the scientific world.

Albeit these questionnaires are completely anonymous, ASI collected consents by students (or their parents/legal guardians for children under 18 years old), so that this work is carried out in accordance with the principles outlined in the ethical policy of this journal.

The paper is organized as follows. An overview of the "Alternanza scuola-lavoro" project, and of the SSDC expertise are given in Section 2. In section 3 we present the SSDC activities proposed to high-school students, focusing on the physics topics involved and describing the hands-on projects. In Section 4, a selection of the questions and the listing of the students' answers (that together compose our reference dataset) are reported, and our data analysis is presented. In Section 5 the results are discussed and some conclusions are drawn.

## 2. OVERVIEW

To put into context the pedagogical experiences here proposed and analysed, it is important to describe both the Italian government program upon which the ASI-SSDC project is based, and the long-term SSDC experience in space data dissemination and outreach.

### 2.1 The "Alternanza scuola-lavoro" program

The "Alternanza scuola-lavoro" program has been established by the Italian law 107/2015, also known as "La Buona Scuola" (i.e. "The Good School"), introducing compulsory periods of workplace training for up to 400 hours per year in all upper secondary level students (final three years). From the May 2018 official report (MIUR, 2018), it results that almost all the Italian schools participated to the program (89% of the total and 94% for public ones), with almost 1 million of students involved during the 2016/2017 school year.

In this context, ASI participates to the "Alternanza scuola-lavoro" initiatives designing specific educational pathways to be followed by the students and publishing open opportunities for the schools. These opportunities are released together with a pedagogical plan and the number of students that can be hosted at the ASI HQ during the activities.

If the number of applying schools is larger than available places, a selection is performed, taking into account the scientific orientation and geographical position of the candidate schools, preferring those more oriented to STEM (Sciences, Technology, Engineering, Mathematic) studies and outside the Lazio region (where the ASI HQ is located).

Two different types of programs are offered by ASI, the "standard" one and the "ad-hoc" one. The "standard" pathway includes well pre-defined stages:

1. the ASI structure is presented to the participating schools, showing the organization, the activities, the peculiarities and the definitions of professional roles;
2. a first stage is then performed at the ASI HQ making the students aware of the different activities that will be proposed;
3. a final, "long", stage contemplates the real working activities of the students inside the ASI HQ, as detailed in the teaching plan.

On the contrary, the "ad-hoc" type has no pre-defined structure, so that it can be best adapted to the particular aim for which is proposed. For the school-year 2018/'19, in addition to the standard one, this kind of ad-hoc program has been proposed for the first time by ASI, developing a theme regarding the mineralogical analysis of planetary surfaces using MATISSE (Multi-purpose Advanced Tool for the Instruments for Solar System Exploration). Since no questionnaire has been given to the participants to this program, it will not be discussed in the results of this work, however some details will be given in subsection 2.4.

**2.2 The SSDC scientific expertise**

The Space Science Data Center of the Italian Space Agency has its roots in the early '90s with the Data Center of the BeppoSax X-ray astronomy satellite, then evolved into the ASI Science Data Center (ASDC), established in 2000 as a multi-mission facility linked to high-energy astrophysics data managing and exploitation. During the latest years, the center expanded its activities to incorporate different branches of space sciences, such as stellar physics, planetary and exoplanetary sciences, changing its name to the current one.

The main tasks of SSDC are the archiving, analysis, distribution of space mission data, and the development of scientific tools to extract high-level scientific information from the data of interest. These goals have been traditionally accomplished by following an "operative approach" involving both scientists and industrial partners. The center is constituted by personnel from the Italian Space Agency, from the National Institute for Astrophysics (INAF), and from the National Institute for Nuclear Physics (INFN), with an Information and Communication Technology (ICT) support provided by industrial partners (that currently are Telespazio and Serco).

This approach allowed for the development of software tools oriented to the real needs of the scientific community of reference. Some of these tools have become milestones for the community, such as the Multi Mission Interactive Archive (MMIA), the SED (Spectral Energy Distribution) Tool and MATISSE.

The center recently expanded its interests to the outreach and educational fields by participating to several local events (e.g., European Researcher Nights and Maker Faire) showing its activities to the general public, and the "Alternanza scuola-lavoro" project proposed by ASI has been immediately viewed as a unique opportunity to demonstrate the capabilities of its analysis tools not only to professional astronomers, but also to light-trained users.

In this context, for the school years 2017/'18 and 2018/'19 the possibility of interacting with scientific tools and data born from the SSDC expertise has been provided by ASI to the students involved in the "Alternanza scuola-lavoro" activities.

## 3. Proposed activities for high-school students

### 3.1 The physics beyond the proposed SSDC activities

During the first stage of the project, a brief introduction in both high-energy astrophysics and solar system exploration topics, which are the subject of the proposed activities, have been given by SSDC researchers in the form of oral presentations of roughly half an hour each. This is fundamental to allow the students to correctly understand the activities to be performed, by also connecting them to curricular activities investigated in class. This part includes personal anecdotes or experiences, and is supported by some PowerPoint slides as visual aid to enhance the audience engagement and understanding. Many of the topics introduced during the first stage at SSDC (such as electromagnetism, flight dynamics, chemistry, and others) may be connected to curricular activities and further investigated in class, also with the help of the provided supporting material (slides, useful links, etc.).

*3.1.1 The Universe seen through gamma-rays*

The students and their accompanying teachers are given a short introduction to the electromagnetic radiation, in particular to gamma rays, the most energetic part of the electromagnetic spectrum, and then introduced to a brief history of gamma-ray astronomy. On Earth, gamma rays are almost exclusively generated by nuclear explosions, but it was not until the 1960s (during the "Cold War") that US defense satellites discovered by chance that some bursts of gamma rays came from deep space and not from the Earth. Luckily for human-life, gamma rays coming from space are mostly absorbed by Earth's atmosphere (energetic radiation can kill living cells), so gamma-ray astronomy could not develop until it was possible to get the detectors in space onboard of dedicated satellites.

High-energy radiation outside Earth is produced by the hottest and most energetic objects in the Universe, such as supernova explosions and their remnants, neutron stars and pulsars, and regions around black holes both of galactic and extragalactic origin. By exploring the Universe at high energies, scientists can study these extreme cosmic accelerators, search for new physics, testing theories and make experiments that are not possible in our Earth laboratories.

In particular the SSDC activities with the students focussed on emissions from "Active Galactic Nuclei" (AGN), a peculiar class of extragalactic energetic astrophysical sources, thought to be powered by accretion onto supermassive black holes, i.e. by giant rings of gas and dust that circle and fall onto the black holes harbored at these galaxy centers.

AGNs produce radiation across the entire electromagnetic spectrum. Material surrounding the black hole forms a rotating disk that gets heated through the viscous dissipation of gravitational energy, and radiates thermal emission mainly at UV and X-ray energies.

The still not well-understood AGN central engine process also leads to the formation of collimated jets of plasma propagating in a direction perpendicular to the accretion disk, as observed in the case of radio loud AGN. In the radio-loud AGN case, the observed non-thermal radiative power of the jets is a significant fraction of the total bolometric luminosity. The jets are launched at relativistic speeds up to distances that can have extensions far beyond the size of the host galaxy, also featuring strings of compact plasmoids, as indicated by several radio observations.

The non-thermal emission observed in relativistic AGN jets is produced by synchrotron and inverse Compton physical processes, and may be modeled by considering the jet particle energy composition and distribution, the magnetic field values, particle propagation, energy absorption and losses. The details of theoretical modeling of AGN emission is still a debated subject which goes beyond the scope of this paper.

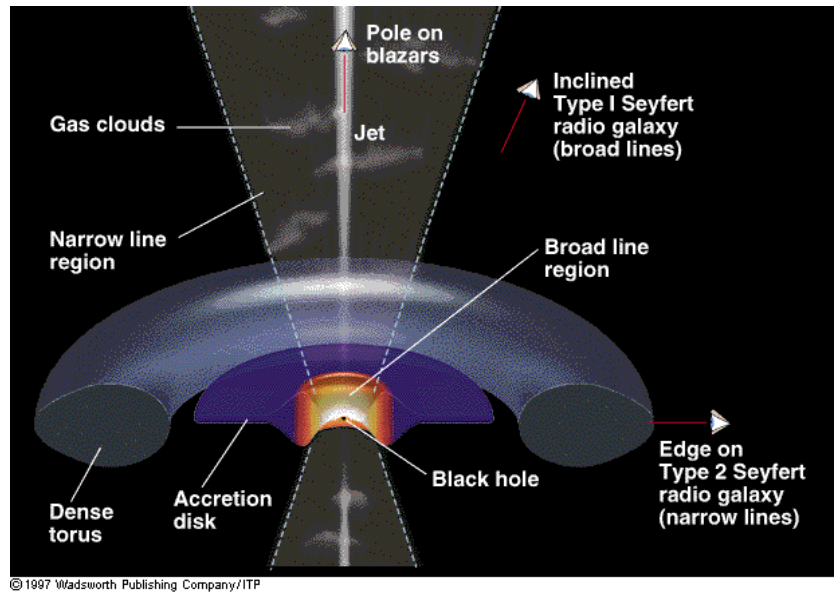

*Figure 1. Illustrative representation of the unified AGN model. Credit: Wadsworth Publishing Company/ITP*

Many observational properties in AGN, though, seem to depend on the viewing angle, which is the main geometrical factor on which the so-called "unified model" of AGNs is based (Antonucci, 1993; Urry & Padovani, 1995). This model unifies all the different types of AGN into a standard model in which the only physical parameter that vary among them is the intrinsic power, while the other observational differences are explained in terms of obscuration and beaming that depend on the angle from where the observer sees the source. Figure 1 shows a schematic view of the AGN unified model. For a recent updated review of AGN classification and observational properties, see for example (Padovani, 2017).

Blazars (blazing quasi-stellar object) are a subclass of radio-loud active galactic nuclei with relativistic jets pointing toward the observer.

The blazar spectral energy distribution (SED) is, in general, characterized by two broad bumps: the low-energy one, spanning from the radio to the X-ray band, is attributed to synchrotron radiation, while the high-energy one, from the X-ray to the $\gamma$-ray band, is thought to be due to inverse Compton (IC) emission. In the leptonic model, this second component is due to relativistic energetic electrons scattering their own synchrotron photons—Synchrotron self-Compton (SSC)— or photons external to the jet—External Compton (EC).

Blazars exhibit episodes of extreme and rapid variability in gamma-rays. The multiwavelength behavior of the source during extreme flares challenges the simple one-zone leptonic theoretical models, and gamma-ray observations of flaring blazars and simultaneous multiwavelength data are thus the key to investigate possible alternative theoretical scenarios.

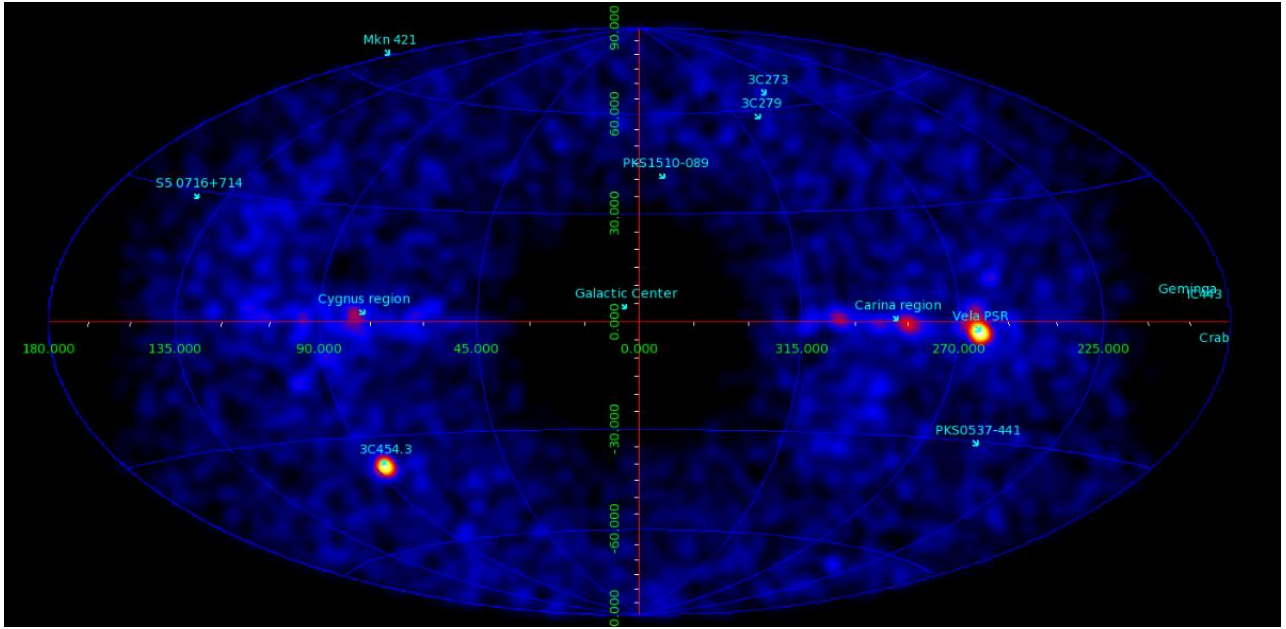

*Figure 2. AGILE all-sky gamma-ray map during the period 22-24 June, 2016, in Aitoff projection and Galactic coordinates, showing the distant source 3C 454.3 in a flaring state.* **Credits: The AGILE Team.**

As an example of extreme flaring episodes, we show in Figure 2 the all-sky gamma-ray photon counts map, in Aitoff projection and Galactic coordinates, as seen by the AGILE gamma-ray imager during the period 22-24 June, 2016. The distant blazar source 3C 454.3 appeared as the brightest source of the entire gamma-ray sky during these days, even brighter than much closer sources within our galaxy, such as the Vela pulsar (Vela PSR in the image), clearly visible on the Galactic plane.

The observation and analysis of real gamma-ray data from these "monster" black holes will be the subject of the student's hands-on activities in the following long stage phase.

*3.1.2 Solar System exploration*

Planetary exploration missions, differently from other kinds of astronomical mission, allow to acquire a wealth of different observations (not only linked to electromagnetic radiation, but also by counting particles, or measuring chemical elements) in the direct proximity of celestial objects.

Since the second half of the past century more than 100 missions allowed us to explore planets, satellites and asteroids by means of fly-bys or from their orbits, providing a totally new perspective of objects until then only studied by Earth-based telescopes.

The solar system activities here proposed to the students were focused on the determination of the composition of the surface of the asteroid Vesta using infrared spectroscopy and, therefore, the students were introduced to the field of research of infrared spectroscopy of rocks and minerals by means of a related presentation.

Atoms inside mineral lattices are not perfectly still, but undergo intra-electronic, vibrational and rotational transitions at energies corresponding to those of the infrared part of the electromagnetic spectrum. Therefore, these minerals absorb (or emit) radiation well constrained by the particular transition involved.

In particular at wavelengths corresponding to the Near Infrared (NIR – i.e., roughly from 1 to 2.5 microns) electronic transition bands (due to transition metals, like $Fe^{2+}$ and $Fe^{3+}$) and vibrational bands (due to water and hydroxyl ion) are present. They are originated either by excited electrons inside *d* and *f* incomplete orbitals, or from transitional metal cations, of which titanium, vanadium, chrome, manganese, iron, cobalt, nickel and copper are crucial for the study of planetary surfaces as they are widely found on the surfaces of rocky celestial objects (Burns, 1993).

Of major interest for these elements at these wavelengths are the crystal field transitions, depending upon symmetry properties of their *3d* orbitals with nearby ligands or anions: in a hypothetical gas constituted by ions all these five orbitals would have the same energy, but inside a crystal lattice this degeneracy is eliminated.

A separation in energy among $e_g$ (i.e., *$d_{z2}$, $d_{x2}$, $d_{y2}$*) and $t_{2g}$ orbitals (i.e., *$d_{xy}$, $d_{yz}$, $d_{zx}$*) will be set, since electrons in the former ones will be repelled more than those in the latter.

Following the CFT (Crystal Field Transitions) model, this energy difference will obey to a "center of gravity" rule: if the energy of the $e_g$ group is raised by $0.6\Delta_0$ from the original energy, then the energy of the $t_{2g}$ group will be lowered by $0.4\Delta_0$.

Another degeneracy decrease is introduced in specific minerals, such as pyroxenes, olivines and hematite, by their not symmetric structure: for the coordination sites of $Fe^{2+}$ in olivines and pyroxenes this makes it possible to differentiate every single mineral by simply looking at the positions of the bands around 1 and 2 microns.

A sensor capable of acquiring and spectral decomposing the (sun) light reflected by a planetary surface would thus allow the identification of the minerals composing it by comparing the detected absorption bands with those retrieved from known minerals analysed in laboratory.

In the case of Vesta, in particular, the imaging spectrometer VIR onboard the NASA Dawn mission has been widely used to characterize the surface during its orbital phase of this asteroid and relate it to a class of meteorites previously found on Earth and tentatively linked to Vesta. These meteorites, called HED (from the three families composing them: Howardites, Eucrites and Diogenites – Binzel et al., 1993) and widely studied in laboratory, show an infrared spectrum up to 2.5 microns very similar to that of the surface of Vesta which is deeply characterized by the presence of olivine minerals, whose, as already stated above, crystal field transitions originate two wide absorption bands near 1 and 2 microns.

In the work by Palomba et al. (2014), used as base for the educational activity here explained, the focus was mainly on the Dark Material Units, areas of Vesta characterized by a surface albedo lower than the surroundings.

Palomba and colleagues exploited a set of spectral parameters related to the olivine absorption bands to accurately characterize 123 Dark Material Units on Vesta, succeeding to extrapolate information about their connection with HED meteorites, their mineralogical composition and the origin of the darkening agents causing the albedo variations and likely due to impacts of carbonaceous chondrites.

**3.2 Hands-on SSDC activities**

Some weeks after the brief stage phase above mentioned, the students came back to ASI HQ for the long stage phase, being involved in the SSDC activities for two days, one devoted to the high-energy astrophysics topic and one to the solar system exploration. In particular, the hands-on activities pointed to replicate scientific analysis already published by professional researchers by making use of real space data.

*3.2.1 Analysing black hole data with the AGILE-LV3 Tool*

The high-energy astrophysics activities aimed to reproduce published scientific results using the interactive online AGILE-LV3 Tool[1] (Pittori, 2019) developed for the gamma-ray AGILE satellite (Tavani et al., 2009) data analysis. The proposed working cases involved the analysis of a couple of exceptional bright flares, such as the one shown in Figure 2, seen in gamma-rays from monster black holes in distant galaxies, reproducing published alert results (e.g., see Astronomer's Telegrams: ATel #9186, Munar-Adrover et al., 2016, and ATel #7631, Lucarelli et al., 2015).

---
[1] http://www.ssdc.asi.it/mmia/index.php?mission=agilelv3mmia

By using the AGILE-LV3 tool, the students, divided in small groups, were asked to:

1. Search the entire public AGILE archive and find data of the selected cosmic sources with enormous black holes at their center, responsible of impressive transient flares of emitted energy, as reported in publications in the form of short Astronomer's Telegrams (ATels), previously provided to each group.
2. Generate with the click of the mouse a so-called "light-curve", that is a graph that shows the brightness of the object over a long period of time (many years), with default time-bin parameters of about 1 month.
3. Identify the month of maximum flux brightness, and perform a refined analysis over shorter time periods, trying to reproduce the results reported in the ATels.
4.

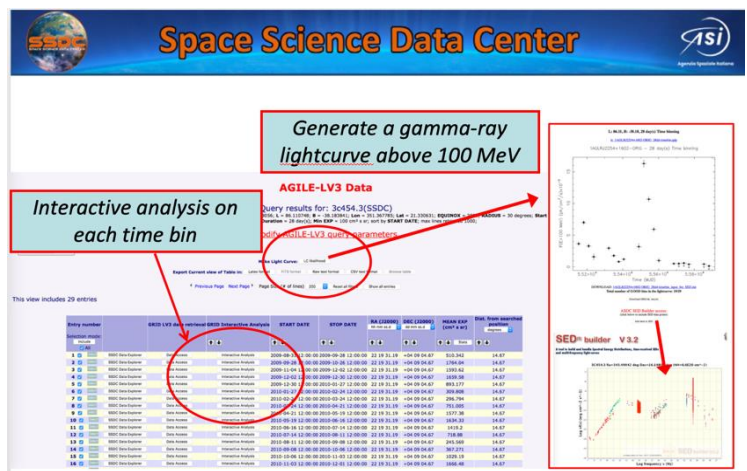

*Figure 3. Combined image for illustrative purposes of the proposed hands-on analysis with the AGILE-LV3 and SED SSDC tools.* **Credits: Screenshots from public online services provided by ASI-SSDC.**

5. Finally, the results obtained by the students with the AGILE-LV3 tool can be easily inserted in the multi-mission SSDC SED Tool, and directly compared with all data available for the selected source from other space missions. The resulting image is a graph of the energy emitted as a function of different wavelengths: from radio, optical, X-rays, to gamma-rays, see also Figure 3.

*3.2.2 Analysing asteroid data with the MATISSE Tool*

As already stated, the solar system activities has been conducted by following the work by Palomba et al. (2014) about the dark regions of the asteroid Vesta. In that work indicative spectral parameters near 1 and 2 microns from Dawn-VIR infrared imaging spectrometer observations have been used to infer the mineralogy, analogous to those of the HED meteorites, and the abundance of darkening agents, likely due to impacts of carbonaceous chondrites, of the surface of Vesta, .

The students used the MATISSE tool[2] (Multi-purpose Advanced Tool for the Instruments for the Solar System Exploration, Zinzi et al., 2016) to find useful Dawn-VIR observations acquired over areas of Vesta specified in the catalogue presented by Palomba et al. (2014).

MATISSE is the webtool developed at SSDC to find, visualize and analyse data from planetary exploration missions directly inside a web browser and with the aid of projection of data on 3D shape of the target. This latter feature makes it extremely valuable primarily for exploration of minor bodies (i.e., asteroids and comets), whose irregular shapes often make it hard the analysis of data projected using "standard" 2D methods (Zinzi et al., 2018).

For this particular activity, students were asked to obtain:

- reflectance at 1.2 microns;
- radiance at 5 microns;
- band depth II;
- band center I;
- band center II;

Apart from the first one, directly obtained by the VIR data, the last three parameters are retrieved thanks to the addition in MATISSE of the spectral parameters' computation by Longobardo et al. (2014). All these parameters are needed to identify and characterize every dark material unit on Vesta, since they are defined not by means of an absolute albedo (at 1.2 microns) value, but from being darker than the surroundings.

This definition is allowed by the use of an imaging spectrometer, which also allows the contemporary analysis of thermal radiation emitted by the target: if an area is darker than the surroundings at 1.2 microns and has a level of radiance at 5 microns not lower than the sorroundings (i.e., is not in shadow), it is identified as a real dark material unit (Figure 4).

---

[2] https://tools.ssdc.asi.it/matisse.jsp

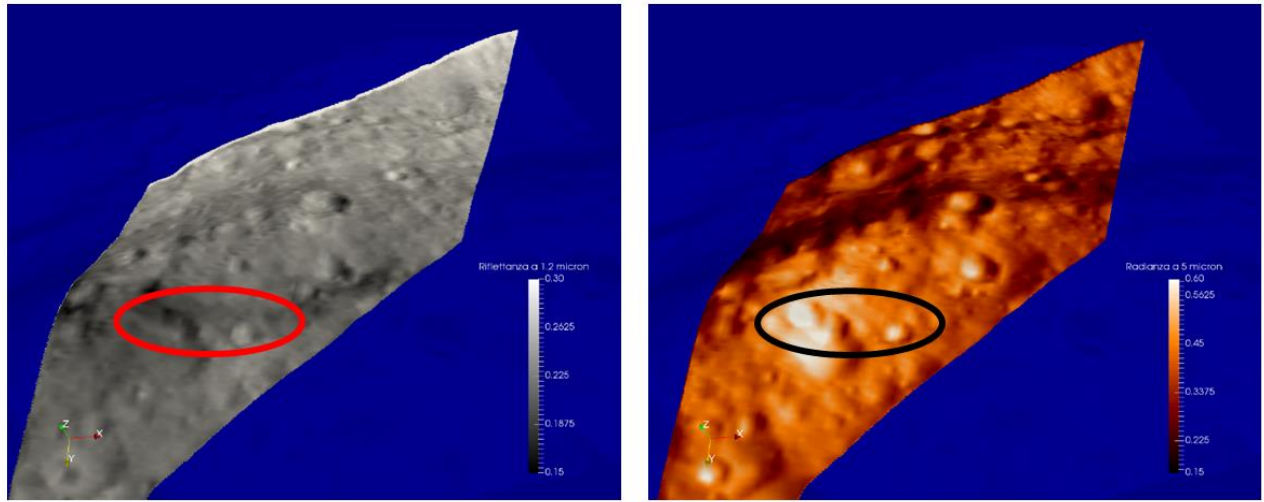

*Figure 4: 3D visualizations of a Dark Material Unit of Vesta with VIR data superimposed generated by MATISSE: on the left 1.2 micron reflectance, on the right 5 micron thermal emission. It is evident that the area marked with the oval is darker than the surroundings at 1.2 micron and brighter at 5 microns, thus satisfying the requirements for Dark Material Units.*

Once this identification is completed, the other parameters under consideration provide the physical characteristics of the studied area: the relation between the band center II and band center I could distinguish among HED types and the relation between reflectance at 1.2 microns and band depth II is indicative of the abundance of darkening agents and of the regolith grain size.

In order to proceed with the activities, we provided the students (divided in groups of no more 5) with a list of VIR observations belonging to some of the 123 dark material units found by Palomba et al. (2014), as performing a geographical query to the public available datasets "VIR IR" and "VIR Spectral Parameters" would have been an unfeasible work for a light-trained student in only a morning activity.

Therefore the students searched for those observations in the cited datasets and, once produced and downloaded all the needed outputs (one for every parameter already mentioned) from MATISSE, they opened the corresponding FITS files by means of the JS9 online FITS reader, to perform a quantitative analysis.

The first step is to locate a dark material unit on the basis of the reflectance at 1.2 microns and the radiance at 5 microns, followed by the computation of the values of parameters in the interesting area of the images, performed by using the "histogram" function of JS9, restricted to a box region.

At this point every group has at least results from one dark material unit, which, following the work by Palomba et al. (2014) will allow to identify its HED classification and darkening agents abundance and the activity could be considered successfully concluded.

In both high-energy astrophysics and solar system exploration cases, students were asked to take note of their activity, and include all their results in a logbook ("Diario di bordo") in the form of a PowerPoint presentation.

The ad-hoc pathway, activated for the school-year 2018/'19, directly descends from this last activity and has been structured in order to make the students better understand both the scientific case proposed, and the MATISSE functionalities.

Differently from the ASI standard programs, the ASI expert involved in the project went 4 times in the selected institute. The first two lessons were more theoretical and concerned the use of infrared spectroscopy in planetary geology, the description of the MATISSE tool, the reading and discussion of the scientific paper focus of the program and a detailed tutorial on how to use MATISSE for the case study.

Then the students, divided in small groups of no more than 5, used MATISSE to analyse the VIR-Dawn data with only a "light" support of the ASI expert and, finally, prepared a PowerPoint presentation about the program under the guide of the ASI expert.

The very last part of the program consisted in a visit to the ASI HQ by the students that, after attending some lectures about the ASI activities, and about the NASA Dawn mission, presented their work to the ASI personnel following their PowerPoint presentation.

It is evident that this kind of more detailed and longer experience involved the students more deeply, leading them to a better understanding of how a research program is conducted.

The same activity has been proposed also for the school-year 2019/2020 and four schools, from different Italian regions have been selected, thus demonstrating the interest of schools to this kind of initiatives. However, due to the COVID-19 emergency, raised in Italy, and then worldwide, since March 2020, the project, ready to be started, has been temporarily suspended, waiting for the evolution of the emergency.

## 4. Student feedback data analysis

**4.1 School Year 2017/2018**

After the completion of all the activities proposed, students were invited to compile an online questionnaire (whose questions and answers are shown in Appendix, Table 1), whose main aim was to understand the impact of the SSDC activities on their perspectives and to find relationships between these responses and the natural attitudes of the students.

There were 20 students involved (12 females and 8 males), equally distributed among the two different schools participating to the activities. The questionnaire was composed by 3 different sections regarding, respectively, "Home and school habits", "Current experience" and "Future perspectives".

*4.1.1 Home and school habits*

In the first section we asked which subjects have been studied at school among biology, chemistry, physics and astronomy, and we subsequently investigated the time spent weekly (at home and school) with computers and smartphone and how much of this time was dedicated to educational purposes.

All the students studied biology, chemistry and physics whereas only half of them (i.e., one school) already studied astronomy.

The second series of questions put in evidence that the 85% of the students uses computer less than 2 hours per week at school, whereas half of them spends more than half an hour daily at the computer at home, and 90% uses the smartphone more than 1 hour per day.

It is then interesting to note that the number of students using the computer at home mainly for educational purposes is exactly equal to that of students not using it with this principal aim.

*4.1.2 Current experience*

The first series of questions posed in this section regards how the tasks to be done have been adequately explained, how much they were interesting and if there were sufficient time to ask questions. The possible answers were "totally disagree", "partially disagree", "don't know", "partially agree", "totally agree" and, by tagging the last two answers as "positive", 95% of the students answered positively to all of these 3 questions.

The next two series of questions are the same, but for high-energy and solar system, respectively:

- Has the team work been useful to resolve problems?

- Was I capable of resolving problems?
- Was the time to pose question enough?
- Did I resolve correctly the task?

Following the same classification as before, answers to these questions were almost totally positive: for the high-energy part we always reached 95% of positive results (with 80% only at the third question), whereas for the solar system part positive answers were, respectively, 85%, 80%, 95% and 80%.

We then asked what is the difficulty level of the experiences, obtaining the following results:

- High-energy: Easy 5%, Right 80%, Hard 15% ("Too easy"/"Too hard" 0%)
- Solar system: Easy 15%, Right 60%, Hard 25% ("Too easy"/"Too hard" 0%)

The following part of the section investigated the level of interest (compared to that before the experience): for high-energy astrophysics the 90% declared to have augmented its interested to the matter, whereas the percentage was 80% for the solar system activity.

We then asked if, after the experience, the students thought to better know high-energy astrophysics and solar system exploration arguments, if they better understood the usefulness of space data and if they were more interested in astronomy.

Even in this case positive answers are largely the most common, i.e., 95% for the first question, 90% for the second question, 100% for the third question and 90% for the last question. It is then interesting to note that, asked about the duration of the experience, half of the sample answered with "right", but 30% of the students judged it too short. To end this section, 95% of the students would recommend this experience to their mates.

*4.1.3 For the future*

This section only comprises one question, with which students were asked about if this kind of experiences should be organized more frequently: 95% of them "totally agree" with this statement and 5% "agree" (i.e., 100% of positive answers).

**4.2 School Year 2018/2019**

The questionnaire prepared for the second school year here discussed is slightly different from the previous one (questions and answers are shown in Appendix, Table 2). It comprises 4 sections, namely, Goals, Teaching paths, Contents and Conclusions and, also in this case, there were 20 students from 2 schools.

*4.2.1 Goals*

Here students were asked about the preliminary information given to them by the school about the motivations about the participation to the ASI project and its goals.
The answers to this section reveal that 60% of the students was not informed about the motivations, whereas 65% of them were aware about the project goals before its start.

*4.2.2 Teaching paths*

Here we asked the students if they attended the whole project and if they appreciate the way the project has been designed, and, for both the questions, all the answers were positive (i.e., 100% of Yes).

*4.2.3 Contents*

This is the widest section of the questionnaire, comprising questions about: 1) how much contents corresponded to the student expectation, 2) how much contents were in line with their school works, 3) how students evaluated contents chosen, 4) how much the duration of the theoretical lectures were appropriate, 5) how much understandable the arguments were exposed.
Finally, we asked if the students thought of having enlarged their knowledge.
It is worthy to note that questions 3, 4 and 5 were two-fold (i.e., 3a, 3b), as they have been asked for high-energy astrophysics and solar system exploration, respectively.
In a scale from 1 (worst) to 6 (best) all the first 5 questions resulted in a mean value of at least 4.65, except for answer 2, whose mean value was 3.75.
For the last question, all of the students involved answered positively.

*4.2.4 Conclusions*

Here we asked the students if the project fulfilled its main goals, if it helped raising new needs and if they are planning to attend university after high-school.

The first question mean value was 4.95 (in the before mentioned scale from 1 to 6), whereas 68% and 100% of the students answered positively to the second the third question, respectively.

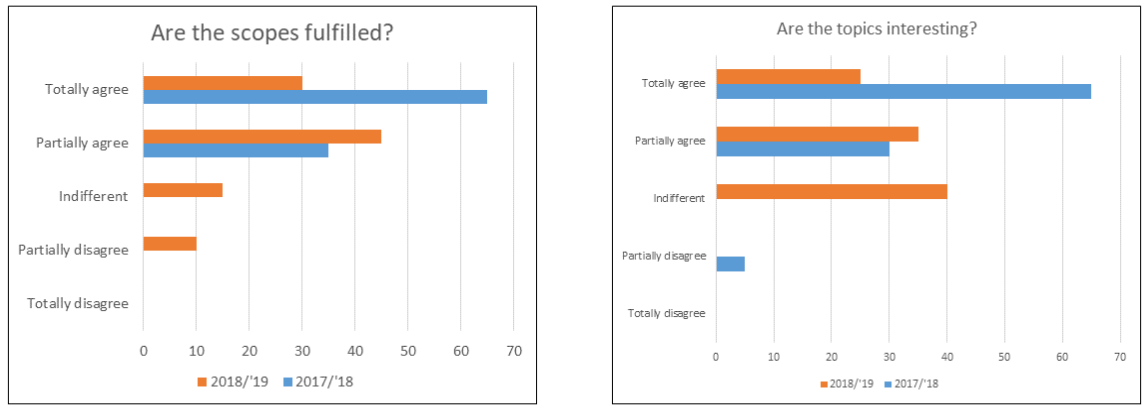

*Figure 5: Histograms displaying how students judged the scopes of the project fulfilled (left) and the topics interesting (right). The values are in percentage of students involved and the colors are different for the two school years analysed.*
*Data on the left were acquired by answering the following questions: "I better understood usefulness of space data tools" (s.y. 2017/'18) and "The main objectives of the project have been fulfilled" (s.y. 2018/'19).*
*Data for the graph on the right were acquired answering to: "I found the topics interesting" (s.y. 2017/'18) and "Topics satisfied my expectations" (s.y. 2018/'19).*

## 5. CONCLUSIONS

The educational projects developed by ASI in the framework of the Italian "Alternanza Scuola-Lavoro" ministerial program are of primary relevance inside the outreach/educational activities of the agency. During the latest years, SSDC actively contributed to these events proposing projects in which the students, by means of the tools developed by the center, can directly use space data to perform real scientific analysis.

In order to examine the effectiveness of the pedagogical approaches applied we distributed questionnaires to the involved students, analysing the answers to highlight their pros and cons and we discussed the results in this work.

These data seem to indicate that the goals of the educational program here discussed have been adequately fulfilled by means of the pedagogical methodology chosen (Figure 5).

The students generally well understood the importance of space data usability and accessibility, and they readily acquired a basic knowledge about the usage of professional tools to search and analyse them for the scientific use cases proposed.

In particular, the answers highlight that the achievement of these goals has been facilitated by the possibility of using the tools soon after the theoretical lectures proposed in which they were first of all introduced to the science topics beyond the hands-on program (i.e., high-energy flares in AGNs and planetary mineralogy) and then made aware about the capabilities of the tools to be used.

Albeit the majority of the involved students judged as balanced the difficulty of the arguments proposed, it is not to be forgotten that a not negligible minority evaluated the lessons as difficult. This could also be due to the fact that the discussed topics are not arguments addressed during school lessons, as evidenced by the results of the second question of the "Contents" section of the 2018/2019 school year. Nonetheless students declared their interest in this type of experiences and would appreciate to be more frequently involved, also with more time available to pose questions and deepen their practice.

Indeed, trying to meet student's suggestions, we also tested a totally new pathway, designed differently from the standard ones here analysed. This ad-hoc pathway exploits a long-lasting interaction between experts and students, by means of several lessons (both theoretical and practical) in the classrooms. This project, although not evaluated here in a quantitative form, has been greatly appreciated by both teachers and students, which participated in a more active and conscious way with respect to those involved in the standard case.

It is also worth noticing that, by spending more time with the students, we succeeded in studying several topics of major interest for the career of a researcher, which could be only marginally discussed during the standard program. These include the structure of a scientific paper, the peer-review process, the difficulty of finding the best observations to characterize a phenomenon, and the understanding of how correctly organize an effective oral presentation about the work.

This is indeed the very final goal of the "Alternanza Scuola-Lavoro", i.e., make students really aware of the kind of jobs they could be employed in the future and, therefore, helping them to choose the best suited studies to get the desired jobs.

# Appendix

|  | < 1 h per week | Between 1 and 2 hours per week | < 30 minutes per day | Between 30 and 60 minutes per day | > 1 hour per day |
|---|---|---|---|---|---|
| How often do you use computer at school for educational purposes? | 9 | 8 | 0 | 0 | 3 |
| How often do you use computer at home? | 1 | 4 | 5 | 4 | 6 |
| How often do you use computer at home for educational purposes? | 3 | 7 | 0 | 5 | 5 |
|  | **Totally disagree** | **Partially disagree** | **Don't know** | **Partially agree** | **Totally agree** |
| Introduction to topics has been appropriate | 0 | 1 | 0 | 8 | 11 |
| I found interesting the topics | 0 | 1 | 0 | 6 | 13 |
| There were enough time to ask questions | 0 | 0 | 1 | 2 | 17 |
| HE –The team work been useful to resolve problems | 0 | 1 | 0 | 6 | 13 |
| HE – I feel myself adequate to solve problems | 0 | 0 | 1 | 5 | 14 |
| HE – There were enough time to ask questions | 0 | 1 | 0 | 6 | 13 |
| HE – I correctly solved the task | 0 | 1 | 0 | 5 | 14 |
| SS – The team work been useful to resolve problems | 0 | 0 | 3 | 2 | 15 |
| SS – I feel myself adequate to solve problems | 0 | 1 | 3 | 8 | 8 |
| SS – There were enough time to ask questions | 0 | 0 | 1 | 4 | 15 |
| SS – I correctly solved the task | 0 | 0 | 4 | 7 | 9 |
| | **Too easy** | **Easy** | **Right** | **Hard** | **Too hard** |
| HE – Activities difficulty level | 0 | 1 | 16 | 3 | 0 |
| SS – Activities difficulty level | 0 | 3 | 12 | 3 | 0 |
| | **Far less than before** | **Less than before** | **As before** | **More than before** | **Far more than before** |
| HE – Interest in the topic | 0 | 1 | 1 | 8 | 10 |
| SS – Interest in the topic | 1 | 0 | 3 | 8 | 8 |
| | **Totally disagree** | **Partially disagree** | **Don't know** | **Partially agree** | **Totally agree** |
| After this experience I know better high-energy | 0 | 0 | 1 | 7 | 12 |
| astrophysics topics | | | | | |
| After this experience I know better solar system exploration topics | 0 | 0 | 2 | 8 | 10 |
| After this experience I understand better space data tools usefulness | 0 | 0 | 0 | 7 | 13 |
| After this experience I am more interested to astronomy | 0 | 0 | 2 | 6 | 12 |
| After this experience I am more interested in working in research in the future | 1 | 1 | 4 | 4 | 10 |
| This experience has been useful and I would recommend it to my peers | 0 | 0 | 1 | 4 | 15 |
| This kind of experience should have been organized more frequently | 0 | 0 | 0 | 1 | 19 |
| | **Too short** | **Short** | **Regular** | **Long** | **Too long** |
| Rate the lenght this experience | 6 | 2 | 10 | 2 | 0 |

*Table 1: Questions posed, with relative possible answers and numbers of them, at the students participating at the 2017/2018 school year activities. Where present, HE stands for High-Energy astrophysics, SS for Solar System Exploration.*

|  | Yes | No |  |  |  |  |
|---|---|---|---|---|---|---|
| Have you been informed about the reasons of this experience? | 8 | 12 |  |  |  |  |
| Have you been informed about the aims of this experience? | 13 | 7 |  |  |  |  |
| Has the educational path been attended entirely? | 20 | 0 |  |  |  |  |
| Did you like the way lectures have been made? | 20 | 0 |  |  |  |  |
| *(1 = very bad, 6 = very good)* | **1** | **2** | **3** | **4** | **5** | **6** |
| Did you know the scopes of the project? | 0 | 2 | 5 | 9 | 4 | 0 |
| Did the contents correspond to you expectations? | 0 | 0 | 0 | 8 | 7 | 5 |
| Do you think educational contents are in line with your school works? | 1 | 1 | 7 | 4 | 7 | 0 |
| HE – How do you evaluate the contents of | 0 | 0 | 3 | 4 | 10 | 3 |

| Question | | | | | | |
|---|---|---|---|---|---|---|
| theoretical lectures? | | | | | | |
| SS – How do you evaluate the contents of theoretical lectures? | 0 | 0 | 2 | 5 | 9 | 4 |
| HE – How do you evaluate the duration of activities? | 0 | 1 | 1 | 3 | 12 | 3 |
| SS – How do you evaluate the duration of activities? | 0 | 1 | 1 | 1 | 13 | 4 |
| HE – how much understandable the arguments were exposed? | 0 | 0 | 4 | 1 | 10 | 5 |
| SS – how much understandable the arguments were exposed? | 0 | 1 | 1 | 3 | 12 | 3 |
| How much do you think the main aims of this projects have been fulfilled? | 0 | 0 | 2 | 3 | 9 | 6 |
| | **Yes** | **No** | | | | |
| Do you think to have enriched your knowledge? | 20 | 0 | | | | |
| Did this path raise new needs in you? | 13 | 7 | | | | |
| Are you planning to attend university? | 20 | 0 | | | | |

*Table 2: Questions posed, with relative possible answers and numbers of them, at the students participating at the 2018/2019 school year activities. Where present, HE stands for High-Energy astrophysics, SS for Solar System Exploration.*